# Feature selection revisited in the single-cell era

Pengyi Yang[1,2,3,*], Hao Huang[1,2], and Chunlei Liu[2]

[1] School of Mathematics and Statistics, University of Sydney, NSW 2006, Australia

[2] Computational Systems Biology Group, Children's Medical Research Institute, University of Sydney, Westmead, NSW 2145, Australia

[3] Charles Perkins Centre, University of Sydney, NSW 2006, Australia

[*] Correspondence: Pengyi Yang (pengyi.yang@sydney.edu.au)

## Abstract

Feature selection techniques are essential for high-dimensional data analysis. In the last two decades, their popularity has been fuelled by the increasing availability of high-throughput biomolecular data where high-dimensionality is a common data property. Recent advances in biotechnologies enable global profiling of various molecular and cellular features at single-cell resolution, resulting in large-scale datasets with increased complexity. These technological developments have led to a resurgence in feature selection research and application in the single-cell field. Here, we revisit feature selection techniques and summarise recent developments. We review their versatile application to a range of single-cell data types including those generated from traditional cytometry and imaging technologies and the latest array of single-cell omics technologies. We highlight some of the challenges and future directions on which feature selection could have a significant impact. Finally, we consider the scalability and make general recommendations on the utility of each type of feature selection method. We hope this review serves as a reference point to stimulate future research and application of feature selection in the single-cell era.

# Introduction

High-throughput biotechnologies are at the centre of modern molecular biology, where typically a sheer number of biomolecules are measured in cells and tissues. While significantly higher coverage of molecules is achieved by high-throughput biotechnologies compared to traditional biochemical assays, the variation in sample quality, reagents and workflow introduce profound technical variation in the data. The high dimensionality, redundancy, and noise commonly found in these large-scale molecular datasets create significant challenges in their analysis and can lead to a reduction in model generalisability and reliability. Feature selection, a class of computational techniques for data analytics and machine learning, is at the forefront in dealing with these challenges and has been an essential driving force in a wide range of bioinformatics applications (1).

Until recently, the global molecular signatures generated from most high-throughput biotechnologies have been the average profiles of mixed populations of cells from tissues, organs, or patients, and feature selection techniques have been predominately applied to such 'bulk' data. However, recent development of technologies that enables the profiling of various molecules (e.g., DNA, RNA, protein) in individual cells at the omics scale has revolutionised our ability to study various molecular programs and cellular processes at the single-cell resolution (2). The accumulation of large-scale and high-dimensional single-cell data has seen renewed interests in developing and need for applying feature selection techniques to such data given their increased scale and complexity compared to their bulk counterparts

To foster research in feature selection in the new era of single-cell sciences, we set out to revisit the feature selection literature, summarise its advancement in the last decade and recent development in the field of deep learning, and review its current applications in various single-cell data types. We then discuss some key challenges and opportunities that we hope would inspire future research and development on this fast-growing interdisciplinary field. Finally, we consider the scalability and applicability of each type of feature selection methods and make general recommendations to their usage.

# Basics of feature selection techniques

Feature selection refers to a class of computational methods where the aim is to select a subset of useful features from the original feature set in a dataset. When dealing with high-dimensional data, feature selection is an effective strategy to reduce the feature dimension and redundancy and can alleviate issues such as model overfitting in downstream analysis. Different from dimension reduction methods (e.g., principal component analysis) where features in a dataset are combined and/or transformed to derive a lower feature dimension, feature selection methods do not alter the original features in the dataset but only identify and select features that satisfy certain pre-defined criteria or optimise certain computational procedures (3). The application of feature selection in bioinformatics is widespread (1). Some of the most popular research directions include selecting genes that can discriminate complex diseases such as cancers from microarray data (4, 5), selecting protein markers that can be used for disease diagnosis and prognostic prediction from mass spectrometry-based proteomics data (6), identifying single nucleotide polymorphisms (SNPs) and their interactions that are associated with specific phenotypes or diseases in genome-wide association studies (GWAS) (7), selecting epigenetic features that mark cancer subtypes (8), and selecting DNA structural properties for predicting genomic regulatory elements (9). Traditionally, feature selection techniques fall into one of the three categories including filters, wrappers, and embedded methods (Fig. 1). In this section, we revisit the key properties and defining characteristics of the three categories of feature selection methods. Please refer to (10) for a comprehensive survey of feature selection methods.

Filter methods typically rank the features based on certain criteria that may facilitate other subsequent analyses (e.g., discriminating samples) and select those that pass a threshold judged by the filtering criteria (Fig. 1a). In bioinformatics applications, commonly used criteria are univariate methods such as *t*-statistics, on which most 'differential expression' (DE) methods for biological data analysis are built (11), and multivariate methods that takes into account relationships among features (12). The main advantages of filter methods lie in their simplicity, requiring less computational resources in general, and ease of applications in practice (13). However, filter methods typically select features independent from the induction algorithms (e.g., classification algorithms) that are applied for downstream analyses, and therefore, the selected features may not be optimal with respect to the induction algorithms in the subsequent applications.

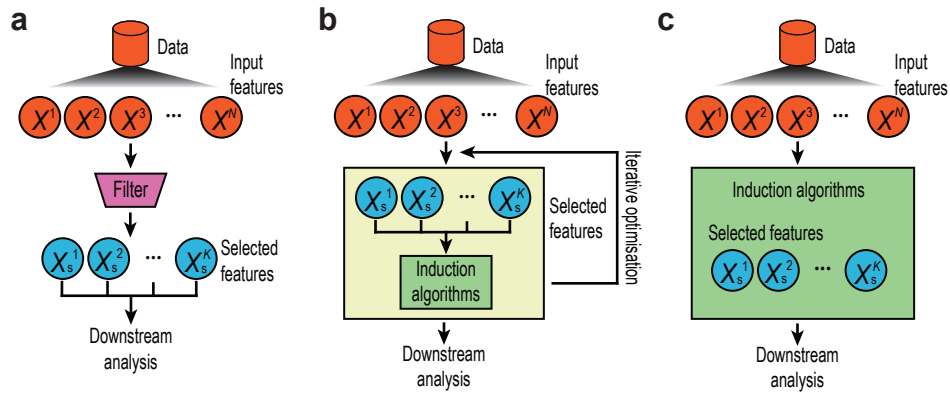

**Figure 1**. Schematic illustrations of typical filter (**a**), wrapper (**b**), and embedded methods (**c**) in feature selection.

In comparison, wrappers utilise the performance of the induction algorithms to guide the feature selection process and therefore may lead to features that are more conducive to the induction algorithm used for optimisation in downstream analyses (14) (Fig. 1b). A key aspect of wrapper methods is the design of the feature optimisation algorithms that maximise the performance of the induction algorithms. Since the feature dimensions are typically very high in bioinformatics applications, exhaustive search is often impractical. To this end, various greedy algorithms, such as forward and backward selection (15), and nature-inspired algorithms, such as the genetic algorithm (GA) (16) and the particles swarm optimisation (PSO) (17), were employed to speed up the optimisation and feature selection processes. Nevertheless, since the induction algorithms are included to iteratively evaluate feature subsets, wrappers are typically computationally intensive compared to filter methods.

While filters and wrappers separate feature selection from downstream analysis, embedded methods typically perform feature selection as part of the induction algorithm itself (18) (Fig. 1c). Akin to wrappers, embedded methods optimise selected features with respect to an induction model and therefore may lead to more suited features for the induction algorithm in subsequent tasks such as sample classification. Since the embedded methods perform feature selection and induction simultaneously, it is also generally more computationally efficient than wrapper methods albeit less so when compared to filter methods (19). Nevertheless, as feature selection is part of the induction algorithm in embedded methods, they are often specific to the algorithmic design and less generic compared to filters and wrappers. Popular choices of embedded methods in bioinformatics applications include tree-based methods (20, 21) and shrinkage-based methods such as LASSO (22).

## Advance of feature selection in the past decade

Besides the astonishing increase in the number of feature selection techniques in the last decade, we have also seen a few notable trends in their development. Here we summarise three aspects that have shown proliferating research in various fields and applications, including bioinformatics.

First, a variety of approaches have been proposed for ensemble feature selection, including those for filters (23, 24), wrappers (25), and embedded methods such as tree-based ensembles (26). Ensemble learning is a well-established approach where instead of building a single model, multiple 'base' models are combined to perform tasks (27). Supervised ensemble classification models are popular among bioinformatics applications (28) and have recently seen their increasing integration with deep learning models (29). Similar to their counterpart in supervised learning, ensemble feature selection methods, typically, relies on either perturbation to the dataset or hyperparameters of the feature selection algorithms for creating 'base selectors' from which the ensemble could be derived (30). Examples include using different subsets of samples for creating multiple filters or using different learning parameters in an induction algorithm of a wrapper method. Key attributes of ensemble feature selection methods are that they generally achieve better generalisability in sample classification (31) and higher reproducibility in feature selection (32, 33). Although these improvements in performance typically come with a cost on computational efficiency, ensemble feature selection methods are increasingly popular given the increasing computational capacity in the last decade and the parallelisation in some of their implementations (34-36).

Second, various hybrid methods have been proposed to combine filters, wrappers, and embedded methods (37). While these methods closely resemble ensemble approaches, they do not rely on data or model perturbations but instead using heterogeneous feature selection algorithms for creating a consensus (38). Typically, these include combining different filter algorithms or different types of feature selection algorithms (e.g., stepwise combination of filter and wrapper). Generally, hybrid methods are motivated by the aim of taking advantage of the strengths of individual methods while alleviating or avoiding their weaknesses (39). For example, in bioinformatics applications, several methods combine filters with wrappers in that filters are first applied to reduce the number of features from high dimension to a moderate number so that wrappers can be employed more efficiently for generating the final set of features (40, 41). As another example, genes selected by various feature selection methods are used for training a set of support vector machines (SVMs) for achieving better classification accuracy using microarray data (42). While many hybrid feature selection algorithms are intuitive and numerous studies have reported favourable results compared to their individual components, a fundamental issue of these methods is their ad-hoc nature, complicating the formal analysis of their underlying properties, such as theoretical algorithmic complexity and scalability.

Third, a recent evolution in feature selection has been its development and implementation using deep learning models. These include models based on perturbation (43, 44), such as randomly excluding features to test their impact on the neural network output, and gradient propagation, where the gradient from the trained neural network is backpropagated to determine the importance of the input feature (45, 46). These deep learning feature selection models share a common concept of "saliency" which was initially designed for interpreting black-box deep neural networks by highlighting input features that are relevant for the prediction of the model (47). Some examples in bioinformatics applications include a deep feature selection model that uses a neural network with a weighted layer to select key input features for the identification and understanding regulatory events (48); and a generative adversarial network approach for identifying genes that are associated with major depressive disorders using gradient-based methods (49). While feature selection methods that are based on deep learning generally require significant more computational resources (e.g., memory) and may be slower than traditional methods (especially when compared to filter methods), their capabilities for identifying complex relationships (e.g., non-linearity, interaction) among features have attracted tremendous attention in recent years.

## Feature selection in the single-cell era

Until recently, the global molecular signatures generated from most biotechnologies are the average profiles from mixed populations of cells, masking the heterogeneity of cell and tissue types, a foundational characteristic of multicellular organisms (50). Breakthroughs in global profiling techniques at the single-cell resolution, such as single-cell RNA-sequencing (scRNA-seq), single-cell Assay for Transposase Accessible Chromatin using sequencing (scATAC-seq) (51) and cellular indexing of transcriptomes and epitopes by sequencing (CITE-seq) (52), have reshaped many of our long-held views on multicellular biological systems. These advances of single-cell technologies create unprecedented opportunities for studying complex biological systems at resolutions that were previously unattainable and have led to renewed interests in feature selection for analysing such data. Below we review some of the latest developments and applications of feature selection across various domains in the single-cell field. Table 1 summarises the methods and their applications with additional details included in Table S1.

**Table 1.** Categorisation of feature selection methods applied to the single-cell field.

|          | Category | Methods        | Transcriptomics | Epigenomics | Surface proteins | Imaging   | Multimodal |
|----------|----------|----------------|-----------------|-------------|------------------|-----------|------------|
| Classic  | Filter   | Univariate     | (53–60)         | (61–64)     |                  | (65, 122) |            |
|          |          | Multivariate   | (66, 67)        |             |                  |           |            |
|          | Wrapper  | Greedy         | (68)            |             | (69, 70)         |           |            |
|          |          | Nature-inspired| (71, 72)        |             | (73)             |           |            |
|          |          | Others         | (74, 75)        |             |                  | (76)      |            |
|          | Embedded | Tree-based     | (77)            | (81)        | (82)             | (83, 84)  |            |
|          |          | Shrinkage      | (78, 79)        | (62, 81)    | (85)             |           | (86)       |
|          |          | Others         | (80)            |             |                  | (83)      |            |
| Advanced | Ensemble |                | (87)            |             |                  |           |            |
|          | Hybrid   |                | (88–91)         | (90)        |                  |           |            |
|          | Deep learning |           | (49, 92)        |             |                  | (93)      |            |

### Feature selection in single-cell transcriptomics

By far, the most widely applied single-cell omics technologies are single-cell transcriptomics (94) made popular by an array of scRNA-seq protocols (95). Given the availability of huge amount of scRNA-seq data and the large number of genes profiled in these datasets, a similar characteristic of their bulk counterparts, most of recent feature selection applications in single-cell transcriptomics have been concentrated on gene selection from scRNA-seq data for various upstream pre-processing and downstream data analyses.

Among these, some of the most popular methods are univariate filters designed for identifying differential distributed genes, including *t*-statistics or ANOVA based DE methods (53, 54) and other statistical approaches such as differential variability (DV) (55) and differential proportion (DP) (56). While differential distribution-based methods can often identify genes that are highly discriminative for downstream analysis, they require labels such as cell types to be pre-defined, limiting their applicability when such information is not available. A less restrictive and widely used alternative approach is to filter for highly variable genes (HVGs), which is implemented in various methods including the popular Seurat package (57). Other methods that do not require label information include SCMarker which relies on testing the number of modalities of each gene through its expression profile (58), M3Drop which models the relationship between mean expression and dropout rate (59), and OGFSC, a variant of HVGs, based on modelling coefficient of variance of genes across cells (60). Many scRNA-seq clustering algorithms also implement HVGs and its variants for gene filtering to improve clustering of cells (96). Besides the above univariate filters, recent research has also explored multivariate approaches. Examples include COMET which relies on a modified hypergeometric test for filtering gene pairs (66), and a multinomial method for gene filtering using the deviance statistic (67).

While filters are the most common options for pre-processing and feature selection from single-cell transcriptomics data, the application of wrapper methods are gaining much attention with a range of approaches built and extend on classic methods with the primary goal of facilitating downstream analyses such as cell type classification. Some examples include the application of classic methods such as greedy-based optimisation of entropy (68), nature-inspired optimisation such as using GA (71, 72), and their hybrid with filters (88-90) or embedded methods (91). More advanced methods include active learning based feature selection using SVM as wrapper (74), and optimisation based on data projection (75). The impact of optimal feature selection using wrapper methods on improving cell type classification is well demonstrated through these studies.

Due to the simplicity in their application, the popularity of embedded methods is growing quickly in the last few years especially in studies that treat feature selection as a key goal in their analyses. These include discovery of the minimum marker gene combinations using tree-based models (77), discriminative learning of DE genes using logistic regression models (78), regulatory gene signature identification using LASSO (79), and marker gene selection based on compressed sensing optimisation (80).

Lastly, several studies have compared the effect of various feature selection methods on clustering of cell types (96) and investigated factors that affect feature selection in cell lineage analysis (97). Together, these studies demonstrate the utility and flexibility of feature selection techniques in a wide range of tasks in single-cell transcriptomic data analyses.

**Feature selection in single-cell epigenomics**

Besides single-cell transcriptomic profiling, another fast-maturing single-cell omics technology is single-cell epigenomics profiling using scATAC-seq (51). In particular, scATAC-seq measures genome-wide chromatin accessibility and therefore can provide clue regarding the activity of epigenomic regulatory elements and their transcription factor binding motifs in single cells. Such data can offer additional information that are not accessible to scRNA-seq technologies, and hence can complement and significantly enrich scRNA-seq data for characterising cell identity and gene regulatory networks (GRNs) in single cells (98). Although most application of feature selection have been on investigating single-cell transcriptomes, recent studies have broadened the view to single-cell epigenomics primarily through their application in scATAC-seq data analysis. These analyses enable us to expand the gene expression analysis to also include regulatory elements such as enhancers and silencers in understanding molecular and cellular processes.

Feature selection methods could be directly applied to scATAC-seq data for identifying differential accessible chromatin regions or one can summarise scATAC-seq data to the gene level using tools such as those reviewed in (99) and then feature selection be performed for selecting 'differentially accessible genes' (DAGs) using such summarised data. For instance, Scasat, a tool for classifying cells using scATAC-seq data, implements both information gain and Fisher exact test for filtering and selecting differential accessible chromatin regions (61). Similarly, scATAC-pro, a pipeline for scATAC-seq analysis at the chromatin level, employs Wilcoxon test as the default for filtering differential accessible chromatin regions, while also implements embedded methods such as logistic regression and negative binomial regression based models as alternative options (62). Another example is SnapATAC (63) which performs differential accessible chromatin analysis using the DE method implemented in edgeR (100). In contrast, Kawaguchi et al. (81) summarised scATAC-seq data to the gene level using SCANPY (101) and performed embedded feature selection using either logistic LASSO or random forests to identify DAGs (81). Muto et al. (64) performed filter-based differential analysis on both chromatin and gene levels based on Cicero estimated gene activity scores (102). Finally, DUBStepR (90), a hybrid approach that combines a correlation-based filter and a regression-based wrapper for gene selection from scRNA-seq data, can also be applied to scATAC-seq data. Collectively, these methods and tools demonstrate the utility and impact of feature selection on scATAC data for cell type identification, motif analysis, regulatory element and gene interaction detection among other applications.

**Feature selection for single-cell surface proteins**

Owing to the recent advancement in flow cytometry and related technologies such as mass cytometry (103, 104), and single-cell multimodal sequencing technologies such as CITE-seq (52), surface proteins of the cells have now also become increasingly accessible at the single-cell resolution.

A key application of feature selection methods to flow and mass cytometry data has been for finding optimal protein markers for cell gating (105). A representative example is GateFinder which implements a random forest-based feature selection procedure for optimising stepwise gating strategies on each given dataset (69). Besides automated gating, several studies have also explored the use of feature selection for improving model performance on sample classification. For example, in their study, Hassan et al. (85) demonstrated the utility of shrinkage-based embedded models for classifying cancer samples. Another application of feature selection techniques was recently demonstrated by Tanhaemami et al. (73) for discovering signatures from label-free single cells. In particular, the authors employed a GA for feature selection and verified its utility on predicting lipid contents in algal cells under different conditions. Together, these studies illustrate the wide applicability of feature selection methods in a wide range of challenges in flow and mass cytometry data analysis.

Recent advancement in single-cell multimodal sequencing technologies such as CITE-seq and other related techniques such as RNA expression and protein sequencing (REAP-seq) (106) has enabled the profiling of both surface proteins and gene expressions at the single-cell level. While still at its infancy, feature selection techniques have already found their use in such data. One example is the application of random forest-based approach for selecting marker proteins that can distinguish closely related cell types profiled using CITE-seq from PBMCs isolated from the blood of healthy human donors (82). Another example is the use of a greedy forward feature selection wrapper that maximises a logistic regression model for identifying surface protein markers for each cell type from a given CITE-seq dataset (70).

**Feature selection in single-cell imaging data**

Other widely accessible data at the single-cell resolution are imaging related data types such as these generated by image cytometry (107) and various single-cell imaging techniques (108). Although, the application of feature selection methods in this domain are very diverse, the following examples provide a snapshot of different types of feature selection techniques used for single-cell imaging data analysis.

To classifying cell states using imaging flow cytometry data, Pischel et al. (65) employed a set of filters, including mutual information maximisation, maximum relevance minimum redundancy, and Fisher score, for feature selection and demonstrated their utility on apoptosis detection. To predict cell cycle phases, Hennig et al. (83) implemented two embedded feature selection techniques, gradient boosting and random forest, for selecting most predictive features from image cytometry data. These implementations are included in the CellProfiler, an open-source software for imaging flow cytometry data analysis. To improve data interpretability of single-cell imaging data, Peralta and Saeys (76) proposed a clustering based method for selecting representative features from each cluster and thus significantly reduces data dimensionality. To classify cell phenotypes, Doan et al. (93) implemented supervised and weakly supervised deep learning models in a framework called Deepometry for feature selection from imaging cytometry data. To classify cells according to their response to insulin stimulation, Norris et al. (84) used a random forest approach for ranking the informativeness of various temporal features extracted from time-course live cell imaging data. Finally, to select spatially variable genes from imaging data generated by multiplexed single-molecule fluorescence in situ hybridization (smFISH), Svensson et al. (122) introduced a model based on Gaussian process regression that decomposes expression and spatial information for gene selection.

**Upcoming domains and future opportunities**

The works reviewed above covers some of the most popular single-cell data types. Nevertheless, the technological advances in the single-cell field are extending our capability at a breakneck speed, enabling many other data modalities (109) as well as the spatial locations (110) of individual cells to be captured in high-throughput. For instance, recent development in single-cell DNA-sequencing provides opportunity to analyse SNPs and copy-number variations (CNVs) in individual cells from cancer and normal tissues (111, 112); and single-cell proteomics seems now on the horizon (113, 114), holding great promises to further transform the single-cell field. Given the high feature-dimensionality of such data (e.g., numbers of SNPs, proteins, and spatial locations), we anticipate feature selection techniques to be readily adopted for these single-cell data types when they become more available.

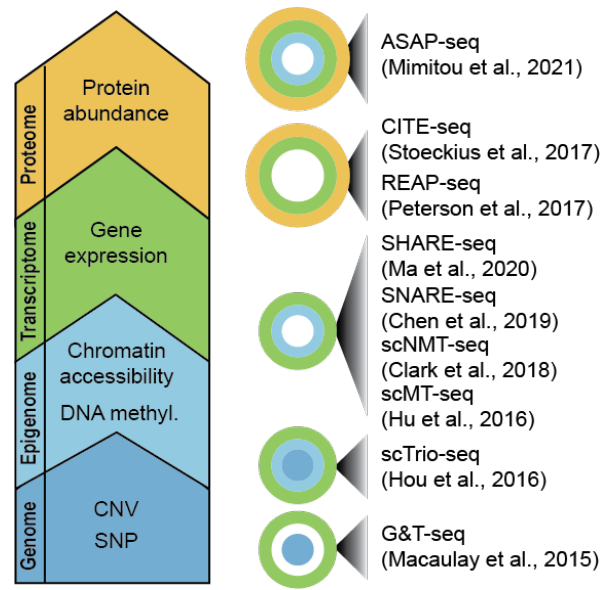

**Figure 2**. A schematic summary of some recent multimodal single-cell omics technologies.

Another fast-growing capability in single-cell field is increasingly towards multimodality. CITE-seq and REAP-seq are examples where both the gene expression and the surface proteins are measured in each individual cell. Nevertheless, many more recent techniques now also enable other combinations of modalities to be profiled at the single-cell level (Fig. 2). Some examples include ASAP-seq for profiling gene expression, chromatin accessibility, and protein levels (115); scMT-seq for profiling gene expression and DNA methylation (116) and its extension, scNMT-seq, for gene expression, chromatin accessibility, and DNA methylation (117); SHARE-seq and SNARE-seq for gene expression and chromatin accessibility (118, 119); scTrio-seq for CNVs, DNA methylation, and gene expression (120); and G&T-seq for genomic DNA and gene expression (121). Given the complexity in data structure in these single-cell multimodal data, feature selection methods that can facilitate integrative analysis of multiple data modalities is in great need. While some preliminary works have emerged recently (86), research on integrative feature selection is still at its infancy and requires significant innovation in their design and implementation.

On the design of feature selection techniques in the single-cell field, most current studies directly use one of the three main types of methods (i.e., filters, wrappers and embedded methods). While we found a small number of them employed hybrid approaches (e.g., [90, 91]), most are relatively straightforward combinations (such as stepwise application of filter and then wrapper methods) as have been used previously for bulk data analyses. The application of ensemble and deep learning based feature selection methods is even sparser in the field. One ensemble feature selection method is EDGE which uses a set of weak learners to vote for important genes from scRNA-seq data (87), and the current literature on deep learning based feature selection in single cells are a study for identifying regulatory modules from scRNA-seq data through autoencoder deconvolution (92); and another for identifying disease associated gene from scRNA-seq data using gradient-based methods (49). Owing to the non-linear nature of the deep learning models, feature selection methods that are based on deep learning are well-suited to learn complex non-linear relationships among features. Given the widespread non-linearity relationships, such as gene-gene and protein-protein interactions, and interactions among genomic regulatory elements and their target genes in biological systems, and hence the data derived from them, we anticipate more research to be conducted on developing and adopting deep learning based feature selection techniques in the single-cell field in the near future.

## Applicability considerations

The works we have reviewed above showcase diverse feature selection strategies and promising future directions in single-cell data analytics. Here, we discuss several key aspects specific to the utility and applicability of feature selection methods with the goal of guiding the choice of methods from each feature selection category for readers who are interested in their application.

**Scalability towards the feature dimension**

A key aspect in the applicability of a feature selection method rests upon its scalability to large datasets. Univariate filter algorithms are probably the most efficient in terms of scalability towards the feature dimension since, in general, the computation time of these algorithms increases linearly with the number of features. We therefore recommend univariate filters as the first choice when working with datasets with very high feature dimensions. In comparison, wrapper algorithms generally do not scale well with respect to the number of features due to their frequent reliance on combinatorial optimisation and therefore will remain applicable to datasets with relatively small number of features. While other factors such as available computational resources and specific algorithm implementations also affect the choice of methods, wrapper algorithms are generally applied to datasets with up to a few hundred features. Embedded methods offer a good trade-off and both tree- and shrinkage-based methods computationally scale well with the number of features (19). Nevertheless, like wrapper methods, embedded methods rely on an induction algorithm for feature selection and therefore are sensitive to model overfitting when dealing with data with small sample size. We recommend choosing embedded methods for datasets with up to a few thousand features when the sample size (e.g., number of cells) is moderate or large. Similarly, hybrid algorithms that combine filter with wrappers or filter with embedded methods also make a useful compromise and can be applied to dataset with relatively high to very high feature dimensions, depending on the reduced feature dimension following the filtering step.

**Scalability towards the sample size**

With the advance of biotechnologies, the number of cells profiled in an experiment is growing exponentially. Hence, apart from the feature dimensionality, the scalability of the feature selection algorithm towards the sample size, typically in terms of the number of cells, is also a central determinant of its applicability to large-scale single-cell datasets. Although classic feature selection algorithms such as filters scale linearly towards the feature dimension, this does not necessarily mean they also scale linearly with the increasing number of cells (53). To this end, the choice is more dependent on the specific implementation of the feature selection algorithms. Methods that purely rely on estimating variabilities (e.g., HVGs) without using cell type labels and fitting models generally scale better due to the extra steps taken by the latter for learning various data characteristics (e.g., zero-inflation). Another aspect to note is the memory usage. Most filter methods require the entire dataset to be loaded into the computer memory before feature selection can be performed. This can be an issue when the size of the dataset exceeds the size of the computer memory. Interestingly, deep learning based feature selection methods could be better suited for analysing datasets with very large number of cells. This is due to the unique characteristic of these methods where the neural network can be trained using small batches of input data sequentially and therefore alleviates the need to load the entire dataset into the computer memory.

**Robustness and interpretability**

Besides algorithm scalability, robustness and interpretability are also important criteria for assessing and selecting feature selection methods. This is especially crucial when the downstream applications are to identify reproducible biomarkers, where the selection of robust and stable features is essential, or to characterise gene regulatory networks, where model interpretability will be highly desirable. A key property of ensemble feature selection methods is their robustness to noise and slight variations in the data, which leads to better reproducibility in selected features (32, 33). We thus recommend exploring ensemble feature selection methods when the task is related to identify reproducible biomarkers such as marker genes for cells of a given type. In terms interpretability, complex models, while often offer better performance in downstream analyses such as cell classification, may not be the most appropriate choices given the difficulties in their model interpretation. To this end, simpler models such as tree-based methods can provide clarity, for example, to how selected features are used to classify a cell and hence can facilitate the characterisation of gene regulatory networks underlying cell identity. Notably, however, significant progress has been made to improve interpretability especially for deep learning models (123). Given the increasing importance in downstream analyses that involves biomarker discovery and pathway/network characterisation in single-cell research, we anticipate increasing efforts to be devoted to improving robustness and interpretability of advanced methods such as deep learning models in feature selection applications.

**Other considerations**

Finally, the choice of feature selection methods also depends on other factors such as programming language, computing platform, parallelisation, and whether they are well documented and easy to use. While most recent methods are implement using popular programming languages such as R and Python which are well supported in various computing platform including Windows, macOS, and Linux/Unix and its variants, their difficulty in application varies and require different levels of expertise from interacting with simple graphical user interface to more complex execution that involves programming (e.g., loading packages in the R programming environment). Methods that optimise for computation speed may use C/C++ as their programming language and may also offer parallelisation. However, these methods are often computing platform-specific and may require more expertise from a specific operating system and programming language from users for their application. Lastly, the quality of the documentation of methods can have a significant impact on their ease of use. Methods that have comprehensive documentations with testable examples could help popularise their application. To this end, methods that are implemented under standardised framework such as Bioconductor (124) generally provide well-documented usages and examples known as "vignette" for supporting users and therefore can be a practical consideration in their choices.

# Conclusions

The explosion of single-cell data in recent years have led to a resurgence in development and application of feature selection techniques for analysing such data. In this review, we revisited and summarised feature selection methods and their key development in the last decade. We then reviewed the recent literature for their applications in the single-cell field, summarising achievements so far and identifying missing aspects in the field. Based on these, we propose several research directions and discuss practical considerations that we hope will spark future research in feature selection and their application in the single-cell era.

**Ethics approval and consent to participate**

Not applicable.

**Consent for publication**

Not applicable.

**Availability of data and material**

Not applicable.

**Competing interests**

The authors declare that they have no competing interests.

**Funding**

P.Y. is supported by a National Health and Medical Research Council Investigator Grant (1173469).

**Authors' contributions**

P.Y. conceptualised this work. All authors reviewed the literature, drafted the manuscript, wrote and edited the manuscript.

**Acknowledgements**

The authors thank the feedback from the members of Sydney Precision Bioinformatics Alliance.